\documentclass[twocolumn,a4paper,superscriptaddress,floatfix]{revtex4}
\usepackage[english]{babel}
\usepackage[utf8]{inputenc}
\usepackage{amsmath}
\usepackage{graphicx,epstopdf}
\usepackage[colorinlistoftodos]{todonotes}
\usepackage{amssymb,epsfig,color,textcase}
\usepackage{hyperref}
\usepackage{doi}

\begin{document}

\title{Orbital Peierls mechanism of the formation of molecular clusters in Na$_2$Ti$_3$Cl$_8$ and in similar systems}

\author{Daniel I. Khomskii}
\affiliation{II. Physikalisches Institut, Universit$\ddot a$t zu K$\ddot o$ln,
Z$\ddot u$lpicher Stra$\ss$e 77, D-50937 K$\ddot o$ln, Germany}

\author{Takashi Mizokawa} 
\affiliation{Department of Applied Physics, Waseda University, Tokyo 169-8555, Japan}

\author{Sergey V.~Streltsov$^*$}
\affiliation{Institute of Metal Physics, S. Kovalevskoy St. 18, 620990 Ekaterinburg, Russia}
\affiliation{Ural Federal University, Mira St. 19, 620002 Ekaterinburg, Russia}
\email{streltsov@imp.uran.ru}

\date{\today}

\begin{abstract}
\end{abstract}

\maketitle
Materials containing well-defined clusters of several atoms - ``molecules in solids'', are very interesting systems. Sometimes these clusters - dimers, trimers or larger clusters - exist just by the very structure of a material. In other cases they are formed spontaneously, as a result of  structural phase transitions   at decreasing temperature. They display very diverse properties, and their origin is an important problem.

Recently an interesting class of such systems attracted big attention. These are layered systems with basically kagom\'e lattice, but with inequivalent, small and large, triangles – ``breathing kagom\'e''.  This is, specifically, Na$_2$Ti$_3$Cl$_8$ \cite{Kelly2019}, but also materials of the class of (Zn,Fe,Ni)$_2$Mo$_3$O$_8$,  which contain (breathing) kagom\'e layers of Mo ions. Recently the structural transition in Na$_2$Ti$_3$Cl$_8$ at 200~K  with the formation of triangular Ti$_3$ clusters\cite{Kelly2019} was studied by ab-initio calculations in \cite{Paul2020}. It was concluded on the basis of the DFT+U results that the usual band effects are not sufficient to explain this transition. The authors then invoked magnetic mechanisms, including complicated  factors such as biquadratic and ring exchanges. In this short paper (the still shorter version was recently published in PRL as a Comment\cite{Khomskii2021}) we want to point out that this overcomplicated explanation is unnecessary, and that the formation of trimers, or breathing kagom\'e lattice in Na$_2$Ti$_3$Cl$_8$,   and in many other systems, has very simple, almost trivial explanation, which exactly follows the line of reasoning first presented in \cite{Khomskii2005a}.

\begin{figure}[b!]
   \centering
   \includegraphics[width=0.49\textwidth]{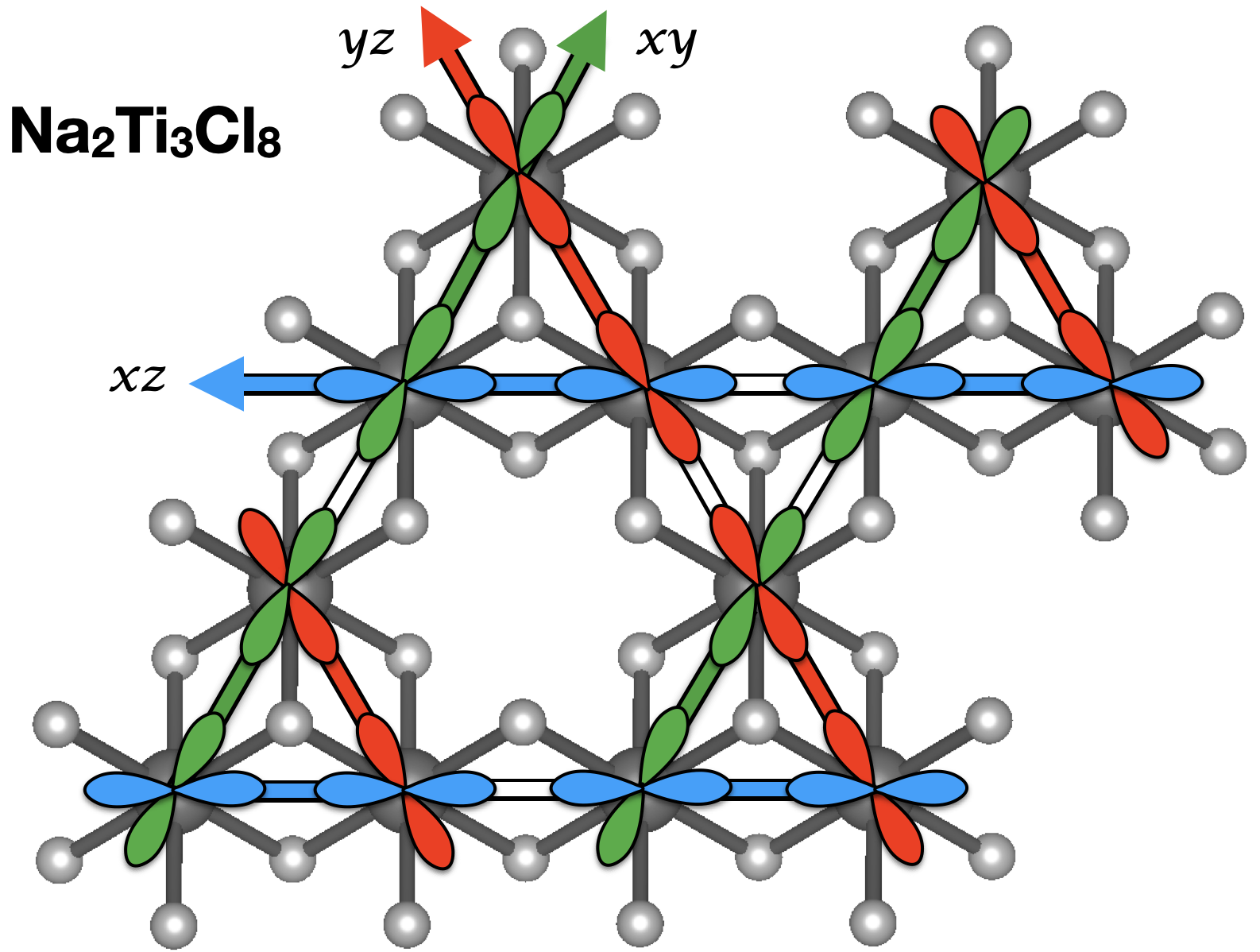}
   \includegraphics[width=0.49\textwidth]{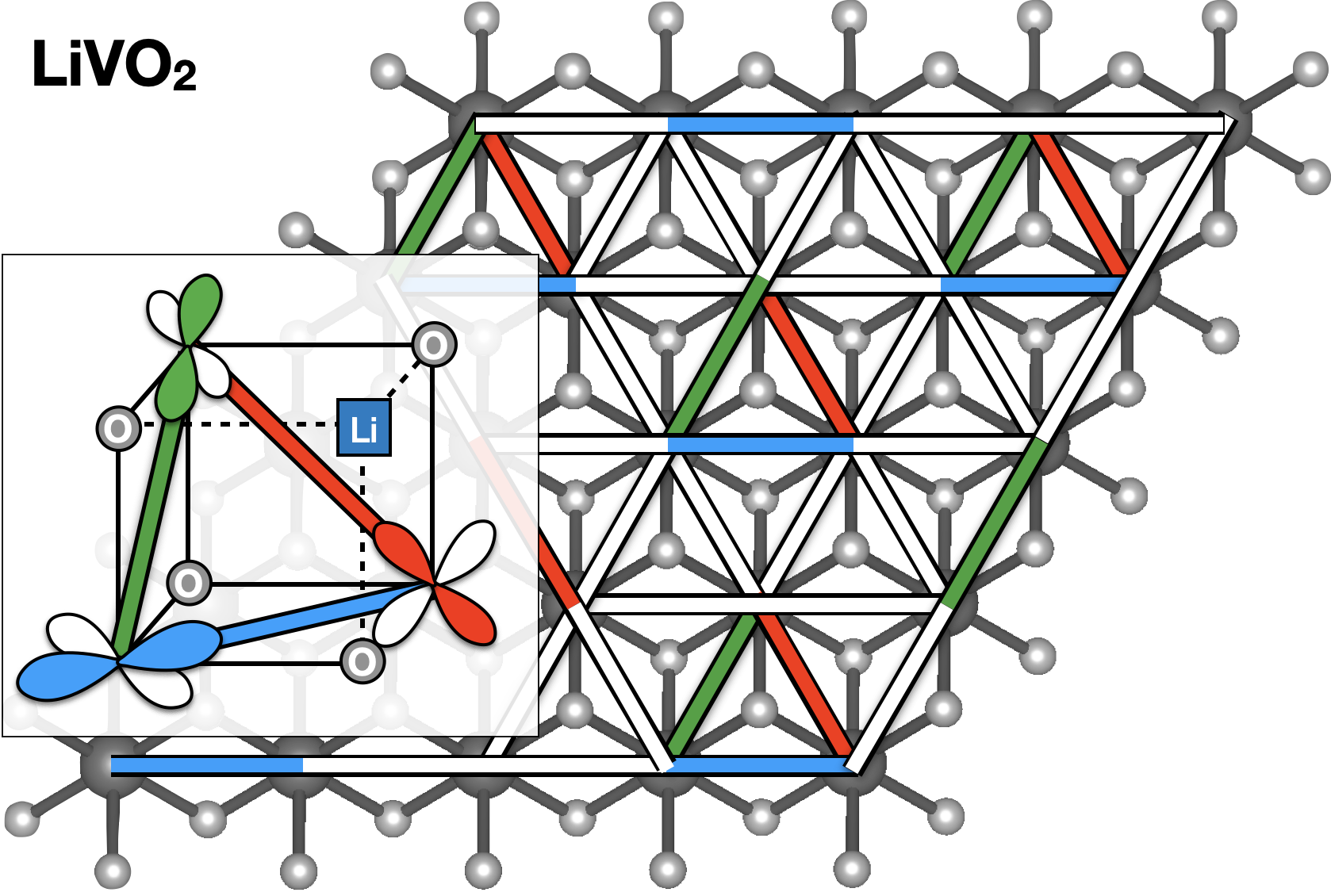}
  \caption{\label{Fig1} Top: Formation of Ti$_3$ trimers in kagom\'e lattice of Na$_2$Ti$_3$Cl$_8$ due to dimerization of chains. Bottom: formation of trimers in triangular lattice of LiVO$_2$ due to the same mechanism: trimerization in 1D chains. Inset shows 3 $t_{2g}$ orbitals (active lobes are colored) direct overlap of which results in formation of $1D$ bands.}
\end{figure}

The simple picture is that of the orbital-driven Peierls distortion.  Ti$^{2+}$ ions have an electronic configuration $d^2$. In the kagom\'e structure of Na$_2$Ti$_3$Cl$_8$ the direct $d-d$ hopping between $t_{2g}$-orbitals would lead to the formation of three one-dimensional ($1D$) bands in three directions, formed by $xy$, $xz$, and $yz$-orbitals, Fig.~\ref{Fig1}.  Each Ti atom lies on the intersection of two such chains and would participate in two bands. With two electrons per Ti, each $1D$ band is half-filled similar to VO$_2$, which would immediately lead to Peierls dimerization. And the breathing kagom\'e structure of Na$_2$Ti$_3$Cl$_8$ corresponds exactly to such dimerization in all three directions! (Fig.~\ref{Fig1}).


Exactly the same mechanism can explain the formation of similar trimers in other family of materials - in Zn$_2$Mo$_3$O$_8$, Fe$_2$Mo$_3$O$_8$ etc, which contain basic kagom\'e layer of Mo ions with the same electronic configuration, Mo$^{4+}$ ($d^2$)~\cite{Khomskii2020}.  In these systems we also have the same dimerization in all three directions, leading to the formation of Mo$_3$ triangular clusters.
It is interesting that the same basic superstructure survives even at some doping, e.g. for LiZn$_2$Mo$_3$O$_8$\cite{Sheckelton2012,Mourigal2014a}.

This ``orbital-driven'' Peierls mechanism can also explain the origin of clusters observed in many other transition metal compounds including not only the formation of dimers (with the resulting beautiful total superstructures) in spinels CuIr$_2$S$_4$ and MgTi$_2$O$_4$\cite{Radaelli2002,Schmidt2004,Khomskii2005a},  but also, e.g., trimerization in another $d^2$ material, LiVO$_2$~\cite{Pen1997}. This system has not kagome, like Na$_2$Ti$_3$Cl$_8$, but triangular lattice. Consequently in this case each V lying on the intersection of three $1D$ chains, participates not in two but in three $1D$ bands, which, with $d^2$ occupation of V$^{3+}$, would give these bands 1/3 filled each. Thus one might expect in this case not Peierls dimerization, but trimerization. And indeed the structure of the low-temperature phase of  LiVO$_2$ shown in Fig.~\ref{Fig1} corresponds exactly to that - it is a trimerization in all three directions. In \cite{Pen1997} this superstructure was explained proceeding from more localised point of view, as a result of a particular orbital ordering. But it is interesting that exactly the same conclusion can be reached also proceeding from the band picture. Also experimentally it was shown \cite{Katayama2009} that the same trimerization occurs also in LiVS$_2$, in which the corresponding transition is really a metal-insulator transition.

\begin{figure}[b!]
   \centering
   \includegraphics[width=0.49\textwidth]{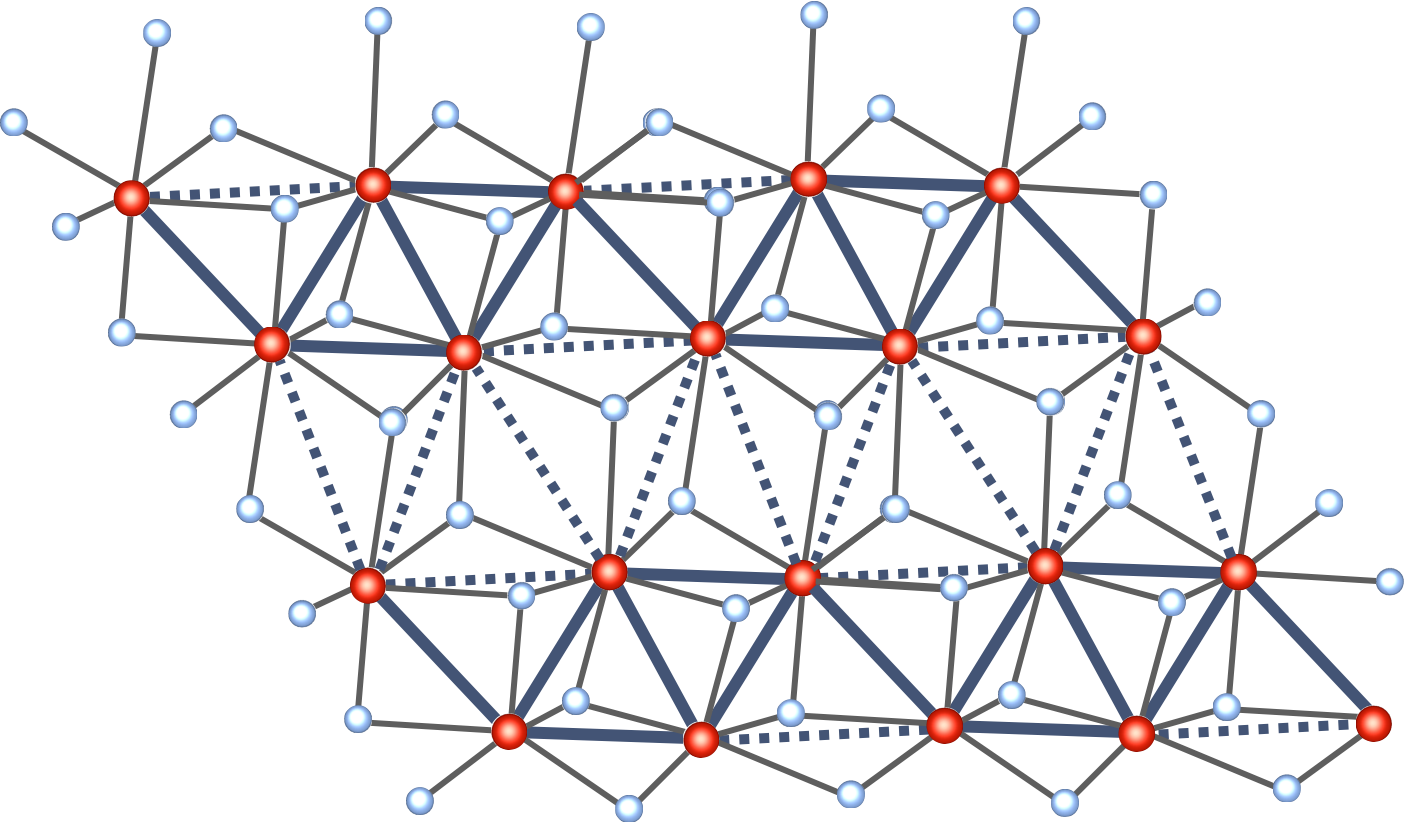}
  \caption{\label{Fig2} ``Diamond necklace'' lattice of ReS$_2$. Re (S) ions are shown by red (blue) balls. Short Re-Re bonds are thick solid blue lines, while long bonds are dashed. Note dimerization in all three straight Re-Re chains.}
\end{figure}

There are also other examples of nontrivial superstructures in solids which can be explained in a similar way. One of such is the `diamond necklace'' lattice in ReS$_2$ and ReSe$_2$\cite{Wilson1969}, see Fig.~\ref{Fig2}. We have a triangular lattice of Re ions and again common edge sharing of ReS$_6$ octahedra. There are three $d$ electrons at each Re and three half-filled ``one-dimensional'' bands, each of them half-filled, which may provide dimerization in all three directions.  This is exactly what is observed experimentally: going along each direction (Re-Re chains) we see dimerization.

Of course, not all superstructures observed in transition metal compounds can be explained by this
picture, especially those with transition metals that are in prismatic coordination rather than in octahedra. In some cases one should rather use more ``chemical'' description, operating with valence bonds formed by available electrons at different orbitals. This is e.g. the situation in AlV$_2$O$_4$, in which such picture gives a natural explanation of the formation of trimers made of V$^{3+}$($d^2$) and of V tetrahedra made by V$^{2+}$($d^3$)\cite{Browne2017,Streltsov2014b}. But it is gratifying that the simple Peierls picture described above gives a good description of the formation of molecular clusters in so many cases.

Interestingly the Peierls-like instability in Na$_2$Ti$_3$Cl$_8$ clearly manifests itself already in the calculations of \cite{Paul2020}. The authors of \cite{Paul2020} focus on the DFT+U results, which do not give lattice instability for the undistorted phase. However, one may notice that the phonon frequencies get imaginary, i.e. the lattice becomes unstable, in simple nonmagnetic DFT calculations of \cite{Paul2020}, see Tab.I in SI\cite{Paul2020}. Thus,  already on the band level the systems tends to distort. This is nothing else but a manifestation of the Peierls instability.  This is a typical situation that including correlations in a simplified way via DFT+U approach\cite{Liechtenstein1995} we often worsen the description of a system tending to form clusters. The famous example is VO$_2$, for which DFT reproduces dimerized phase observed experimentally\cite{Wentzcovitch1994}, while DFT+U stabilizes equally spaced V ions\cite{Wickramaratne2019}. The tendency to form molecular orbitals, which is well described by a kinetic energy in DFT and which is of order of hopping parameter $t$, is substituted in DFT+U by a strong exchange coupling, which is typically much weaker $\sim t^2/U$. DFT+U (in contrast to cluster-DMFT\cite{Biroli2002}) was constructed in a way to localize electrons not on molecular bonds, but on atomic orbitals and therefore it is not surprising that for Na$_2$Ti$_3$Cl$_8$ this method suppressed tendency to formation of trimers due to dimerization in Ti chains.~\cite{Kelly2019}.

Summarising, we see that in quite a few cases the Peierls physics, with orbital structure taken into account,  can successfully explain the formation of ``molecular clusters'' in correlated solids, and this simple physics can be used to rationalise the results of many experiments on many systems.

\section*{Acknowledgements}
We acknowledge German Research Foundation (277146847-CRC-1238), JSPS (19H01853), Russian Foundation for Basic Researches (RFBR 20-32-70019), the Russian Ministry of Education and Science (program ``Quantum''  AAAA-A18-118020190095-4) for financial support.

\bibliography{../../library}

\end{document}